\documentclass[11pt,draftcls,onecolumn]{IEEEtran}
\ifCLASSINFOpdf
\else
  \usepackage[dvips]{graphicx}
  \DeclareGraphicsExtensions{.eps}
\fi
\usepackage[cmex10,tbtags]{amsmath}
\usepackage{amssymb}

\usepackage{algorithmic}
\usepackage{fixltx2e}

\renewcommand{\(}{\left(}
\renewcommand{\)}{\right)}
\renewcommand{\[}{\left[}
\renewcommand{\]}{\right]}

\newcommand{\defeq}{\stackrel{\Delta}{=}}

\DeclareMathOperator{\psinc}{psinc}

\DeclareMathOperator*{\argmax}{arg\,max}
\DeclareMathOperator*{\argmin}{arg\,min}

\ifdefined \algorithmicreturn
\else
	\newcommand{\algorithmicreturn}{\textbf{return}}
\fi
\ifdefined \RETURN
\else
	\newcommand{\RETURN}{\STATE\algorithmicreturn\ \ }
\fi

\begin{document}
%
\title{Nonlinear Digital Post-Processing to\\Mitigate Jitter in Sampling}
%
%
%

\author{Daniel~S.~Weller*,~\IEEEmembership{Student Member,~IEEE,}
        and~Vivek~K~Goyal,~\IEEEmembership{Senior Member,~IEEE}
\thanks{This work was supported in part by the Office of Naval Research through a National Defense Science and Engineering Graduate (NDSEG) fellowship, NSF CAREER Grant CCF-0643836, and Analog Devices, Inc.}
\thanks{D. S. Weller is with the Massachusetts Institute of Technology, Room 36-680, 77 Massachusetts Avenue, Cambridge, MA 02139 USA (phone: +1.617.324.5647; fax: +1.617.324.4290; email: dweller@mit.edu), and V. K. Goyal is with the Massachusetts Institute of Technology, Room 36-690, 77 Massachusetts Avenue, Cambridge, MA, 02139 USA (phone: +1.617.324.0367; fax: +1.617.324.4290; e-mail: vgoyal@mit.edu).}}

\maketitle

\begin{abstract}
This paper describes several new algorithms for estimating the parameters of a periodic bandlimited signal from samples corrupted by jitter (timing noise) and additive noise. Both classical (non-random) and Bayesian formulations are considered: an Expectation-Maximization (EM) algorithm is developed to compute the maximum likelihood (ML) estimator for the classical estimation framework, and two Gibbs samplers are proposed to approximate the Bayes least squares (BLS) estimate for parameters independently distributed according to a uniform prior. Simulations are performed to demonstrate the significant performance improvement achievable using these algorithms as compared to linear estimators. The ML estimator is also compared to the Cram\'{e}r-Rao lower bound to determine the range of jitter for which the estimator is approximately efficient. These simulations provide evidence that the nonlinear algorithms derived here can tolerate 1.4--2 times more jitter than linear estimators, reducing on-chip ADC power consumption by 50--75 percent.
\end{abstract}


 \begin{center} \bfseries EDICS Category: DSP-RECO, SSP-PARE \end{center}
%
\IEEEpeerreviewmaketitle

\section{Introduction}\label{sec:intro}

\IEEEPARstart{A}{nalog-to-digital} converters (ADCs) are affected by different types of uncertainty that introduce noise into the samples of a signal. Two important causes of this noise are inaccurate sample times and variation in the measured signal amplitude. Many well-known approaches exist for separating noise due to amplitude error from the signal. Similarly, noise in the sample times, known as jitter, is also a well-studied phenomenon. However, the effect of jitter is assumed to be negligible in practice because of the low phase noise sampling circuitry employed in conventional ADCs. 

\subsection{Motivation}

The increasing demand for ultra low power ADCs has renewed interest in mitigating jitter using digital post-processing techniques. Because the digital portion of mixed-signal systems like digital sensors and wireless receivers continues to shrink, the analog portion, including the clock generator for the ADC, dominates the size and power consumption of the chip. Such systems would benefit greatly from smaller, more power-efficient analog circuitry; however, the use of such circuitry, with lower voltages, significantly increases noise, such as jitter in the clock signal. The power consumed by an ADC is proportional to the desired accuracy and sampling rate~\cite{Lee00}. In~\cite{Uyttenhove01}, the speed-power-accuracy tradeoff for high-rate ADCs is shown to satisfy
\begin{equation}
\frac{\text{Speed} \times (\text{Accuracy (rms)})^{2}}{\text{Power}} \approx \text{constant}.\label{eq:speedpoweraccuracytradeoff}
\end{equation}

Both~\cite{Brannon00} and~\cite{Walden99} demonstrate that doubling the standard deviation of the jitter cuts down the effective number of bits ($\text{accuracy (rms)} = 2^{\text{ENOB}}$) by one, so the power consumption would need to increase by a factor of four to achieve the same accuracy as before. Clearly, the impact of jitter is not negligible. Incorporating a post-processing method that can recover some of this loss in accuracy should drastically reduce the power consumption of most high-speed ADCs. In addition, the ability to tolerate larger quantities of jitter allows the use of frequency-modulated clocks, which may yield additional benefits like lower electromagnetic interference and radiation~\cite{Lee02}.

\subsection{Problem Formulation}

In this article, nonlinear algorithms for improved classical and Bayesian estimation are developed to recover the parameters of a bandlimited signal. We restrict our attention to signals with a finite basis, leaving the development of streaming algorithms for the infinite-dimensional case a path for further investigation. In particular, we consider a finite segment of a periodic bandlimited signal with period $K$; the chosen basis is one of a periodic sinc function $\psinc_{K}(t) \defeq \frac{\sin(\pi t)}{K\sin(\pi t/K)}$ and its integer shifts. One could also construct a suitable example by applying a smooth windowing function, rather than assuming periodicity. These algorithms are applicable for other signal classes, like splines. For this work, assume the Nyquist frequency is normalized to $\pi$, and the signal $x(t)$ is sampled with an oversampling factor of $M$. The $n$th sample, $y_{n}$, is described by the observation model:
\begin{equation}
y_{n} = \sum_{k=0}^{K-1} \psinc_{K}\(\frac{n}{M} + z_{n} - k\)x_{k} + w_{n}.\label{eq:obsmodelnoisy}
\end{equation}

Let the jitter $z_{n}$ be zero-mean white Gaussian noise, with variance $\sigma_{z}^{2}$; note that the jitter is normalized relative to the critical sampling period. The dominant source of additive noise is assumed to be a signal-independent external noise source (not quantization noise); this additive noise, denoted by $w_{n}$, is also assumed to be zero-mean white Gaussian noise, with variance $\sigma_{w}^{2}$; the additive noise is normalized to the scale of the parameters $x_{k}$. All the noise sources are independent, and they are independent of the input signal parameters $x_{k}$. Define $[\mathbf{H}(\mathbf{z})]_{n,k} = \psinc_{K}(n/M+z_{n}-k)$; then,
\begin{equation}
\mathbf{y} = \mathbf{H}(\mathbf{z})\mathbf{x} + \mathbf{w},\label{eq:obsmodel}
\end{equation}
where $\mathbf{y} = [y_{0},\ldots,y_{N-1}]$, $\mathbf{x} = [x_{0},\ldots,x_{K-1}]$, $\mathbf{z} = [z_{0},\ldots,z_{N-1}]$, and $\mathbf{w} = [w_{0},\ldots,w_{N-1}]$. To emphasize the generality of the model, derivations in the following sections are in terms of $\mathbf{H}(\mathbf{z})$. The $\psinc$ function is used in experiments only for the sake of example.

To keep notation compact, denote by $p(\mathbf{x})$, $p(\mathbf{y};\mathbf{x})$, and $p(\mathbf{y}\mid \mathbf{x})$, the probability density function (pdf) of $\mathbf{x}$, the pdf of $\mathbf{y}$ parameterized by the deterministic variable $\mathbf{x}$, and the pdf of $\mathbf{y}$ conditioned on the random variable $\mathbf{x}$, respectively. In this article, the random variable(s) associated with a pdf are given explicitly only when necessary to avoid ambiguity. Denote the $d$-dimensional multivariate uniform and normal distributions by
\begin{equation}
U_{\mathbf{x}}(\mathbf{a},\mathbf{b}) \defeq \begin{cases}\prod_{i=1}^{d}\(\frac{1}{b_{i}-a_{i}}\)&a_{i} \leq x_{i} \leq b_{i},\ i=1,\ldots,d,\\0&\text{otherwise},\end{cases}\label{eq:uniformdist}
\end{equation}
and
\begin{equation}
\mathcal{N}_{\mathbf{x}}(\mathbf{\mu},\mathbf{\Lambda}) \defeq |2\pi\mathbf{\Lambda}|^{-1/2}\exp\{-\frac{1}{2}(\mathbf{x}-\mathbf{\mu})^{T}\mathbf{\Lambda}^{-1}(\mathbf{x}-\mathbf{\mu})\}.\label{eq:normaldist}
\end{equation}
Then, the likelihood function corresponding to the observation model is
\begin{equation}
\left.\begin{array}{ll}\text{classical:}&p(\mathbf{y};\mathbf{x})\\\text{Bayesian:}&p(\mathbf{y}\mid \mathbf{x})\end{array}\right\} = \int \mathcal{N}_{\mathbf{y}}(\mathbf{H}(\mathbf{z})\mathbf{x},\sigma_{w}^{2}\mathbf{I})\mathcal{N}_{\mathbf{z}}(\mathbf{0},\sigma_{z}^{2}\mathbf{I})\,d\mathbf{z}.\label{eq:likelihood}
\end{equation}
The classical estimation problem is to find the estimator $\mathbf{\hat{x}}(\mathbf{y})$ that minimizes the mean squared error (MSE), knowing only the samples $\mathbf{y}$ and the likelihood function:
\begin{equation}
\mathbf{\hat{x}}(\mathbf{y}) = \argmin_{\mathbf{f}(\cdot)} \mathbb{E}_{\mathbf{Y}}[\|\mathbf{f}(\mathbf{y}) - \mathbf{x}\|_{2}^{2}],\quad \forall \mathbf{x}.\label{eq:nrestproblem}
\end{equation}
Computing this minimum MSE estimator (if such an estimator actually exists) is rarely straightforward; the maximum likelihood (ML) estimator is easier to approximate and will be used instead. The Bayesian estimation problem has the same objective, but with the additional knowledge of a prior on $\mathbf{x}$:
\begin{equation}
\mathbf{\hat{x}}(\mathbf{y}) = \argmin_{\mathbf{f}(\cdot)} \mathbb{E}_{\mathbf{X},\mathbf{Y}}[\|\mathbf{f}(\mathbf{y}) - \mathbf{x}\|_{2}^{2}].\label{eq:bayesestproblem}
\end{equation}
The Bayes least squares (BLS) estimator, $\mathbb{E}[\mathbf{x} \mid \mathbf{y}]$, is the solution to this minimization problem. However, neither the posterior density $p(\mathbf{x} \mid \mathbf{y})$ nor its mean have closed forms for the observation model in~\eqref{eq:obsmodel}.

\subsection{Related Work}

The early literature focuses on linear reconstruction for signals affected by different types of random jitter: \cite{Balakrishnan62} develops a linear interpolation filter for jittered signals, and~\cite{Brown63} explores the effects of sampling and reconstruction jitter. The sampling error due to jitter is examined in~\cite{Liu65}: using low pass interpolation, the MSE is approximately linear in $\sigma_{z}^{2}$. In particular, when the jitter is Gaussian and small enough, and the input power spectral density $S_{xx}(j\Omega) = \frac{1}{2\Omega_{B}}$ is flat, the MSE is approximately $\frac{1}{3}\Omega_{B}^{2}\sigma_{z}^{2}$.

However, these works provide limited insight into the effectiveness of general nonlinear techniques. More recently, \cite{Cox93} provides an iterative weighted least squares fitting algorithm for computing a cubic spline approximation to a jittered input signal. However, the weighted least squares fit uses only a second-order approximation of the input signal. More along the lines of the contributions herein, \cite{Divi04} constructs an Expectation-Maximization (EM) algorithm to estimate a signal sampled by time-interleaved ADCs with timing offsets; however, the offset for each ADC stays constant over time, unlike jitter. Also, a Metropolis-Hastings algorithm is derived in~\cite{Andrieu96} to estimate each sample's timing error and the jitter variance from a sequence of samples; this approach is similar to the Gibbs sampler used in this work.

Another related work, \cite{Marziliano00}, implements several reconstruction methods for finite-length discrete-time signals with sparse coefficients; however, since the jitter is constrained to integer locations to preserve the structure of the DFT, combinatorial methods are employed in lieu of more general statistical approaches. In the context of mitigating read-in and write-out jitter in data storage, a MAP-based estimator is proposed in~\cite{Zhang07} for jitter generated by a first-order Markov process and the bits corrupted by such jitter.

\subsection{Outline}

In Section~\ref{sec:background}, background on Gauss-Hermite quadrature and Monte Carlo methods is provided. For the classical estimation problem, a numerical method is developed in Section~\ref{sec:classical} to approximate the Cram\'{e}r-Rao lower bound, and an Expectation-Maximization (EM) algorithm is presented to compute the ML estimate. Two Gibbs samplers are proposed in Section~\ref{sec:bayesian} for the Bayesian case, one using rejection sampling, and the other using slice sampling. In Section~\ref{sec:results}, simulation results are presented for all these algorithms, and conclusions and future directions for research are discussed in Section~\ref{sec:conclusion}. The problem proposed here, along with further background material, is also discussed in detail in~\cite{WellerThesis}.

\section{Background}\label{sec:background}

The likelihood function described in~\eqref{eq:likelihood} does not have a closed form, and neither does the posterior density. As a result, expressions for the ML estimator, the Cram\'{e}r-Rao bound, and the BLS estimator all involve integrals or expectations that cannot be evaluated directly. More generally, consider the problem of computing $\mathbb{E}[f(x)]$. Two approaches are considered: (a) numeric integration, via Gauss quadrature; and (b) stochastic approximation, via rejection, Gibbs, and slice sampling.

\subsection{Numerical Integration}

Gauss quadrature approximates the univariate expectation $\int \mathbf{f}(\mathbf{x})w(\mathbf{x})\,d\mathbf{x}$ by the sum $\sum_{i=1}^{n} w_{i}\mathbf{f}(\mathbf{x}_{i})$, where $\mathbf{x}_{i}$ and $w_{i}$ are fixed abscissas and weights based on the weighting function $w(x)$, which is the pdf of $X$. The abscissas and weights can be precomputed for a choice of $w(x)$ and $n$ using the eigenvalue-based method in~\cite{Golub69}. Quadrature generalizes to multiple input-multiple output functions $\mathbf{f}(\mathbf{x})$, but if the integrand is not separable, the computational complexity scales exponentially with the number of components in $\mathbf{x}$.

One weighting function of particular interest is the pdf of the standard normal distribution. The related method of quadrature is known as Gauss-Hermite quadrature, since the abscissas and weights are derived from Hermite polynomials. Adapting the abscissas and weights for any univariate normal distribution is easy:
\begin{equation}
\frac{1}{\sqrt{2\pi}\sigma}\int_{-\infty}^{\infty} f(x)\mathcal{N}_{x}(\mu,\sigma^{2})\,dx \approx \sum_{i=1}^{n}w_{i}f(\sigma x_{i} + \mu),\label{eq:gausshermitequad}
\end{equation}
where $x_{i}$ and $w_{i}$ are the abscissas and weights for the standard Normal weighting function. As discussed in~\cite{Davis84}, the error for Gauss-Hermite quadrature goes to zero exponentially fast as long as the $n$th derivative does not increase too fast with $n$; a sufficient condition for convergence is that $f(x)$ does not increase more rapidly than $e^{x^{2}}$.

\subsection{Stochastic Approximation}

Stochastic approximation reduces the problem of calculating an expectation $\mathbb{E}[f(x)]$ to sampling from the distribution $p(x)$ using methods like rejection sampling, Gibbs sampling, and slice sampling.

\subsubsection{Rejection Sampling} Rejection sampling generates samples from a proposal distribution $q(x)$ that envelopes the target distribution $p(x)$; i.e. there exists $c > 1$ such that $cq(x) > p(x)$. Accepted samples have the same distribution as $p(x)$.

If the normalizing constant for a target distribution is unknown, rejection sampling is still effective. Suppose only the form $\tilde{p}(x)$ is known, so $P = \int \tilde{p}(x)\,dx$ is unknown (but assumed to be finite). Then, we choose $c$ such that $cq(x) > \tilde{p}(x)$, and rejection sampling works as before.

One disadvantage of rejection sampling is that if the proposal distribution is a poor approximation to the target distribution, the algorithm generally will reject a prohibitively large number of samples. The ``envelope'' requirement may make choosing a good proposal distribution very difficult. The expected number of iterations between accepted samples is $c/P$. The Gibbs and slice samplers overcome this issue by removing the enveloping requirement.

\subsubsection{Gibbs Sampling} 
\begin{figure}[!t]
\centering
\includegraphics[width=3in]{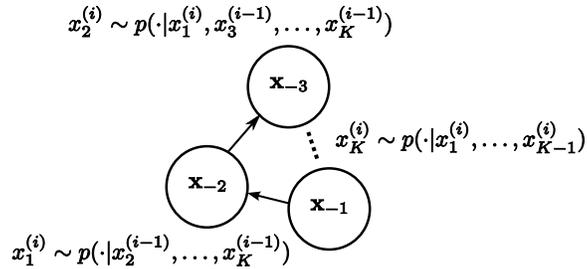}
\caption{Markov chain of the Gibbs sampler. Gibbs sampling consists of repeatedly sampling from these transition distributions in turns until the chain converges to the steady-state. Since the steady-state distribution is the joint pdf $p(\mathbf{x})$, further samples from the transition probabilities will be generated as if they were generated by the joint distribution itself.}
\label{fig:gibbssampler}
\end{figure}

The Gibbs sampler is a Markov chain Monte Carlo method introduced in~\cite{Geman84}. The desired sampling distribution $p(x)$ is the stationary distribution of the constructed Markov chain. Consider the joint probability density function $p(x_{1},x_{2},\ldots,x_{K})$, and define a chain of random variables $x_{k\mid -k} \sim p(x_{k}\mid x_{1},\ldots,x_{k-1},x_{k+1}\ldots,x_{K})$. By augmenting these variables with the last $K-2$ states to become $\mathbf{x}_{-2},\ldots,\mathbf{x}_{-K},\mathbf{x}_{-1}$, the chain becomes a Markov chain; then, $p(\mathbf{x})$ is the stationary distribution of this chain (see Figure~\ref{fig:gibbssampler}).

Sufficient conditions for convergence of the underlying Markov chain are outlined in~\cite{Gelfand00} and~\cite{Casella92}. According to~\cite{Smith93}, separating correlated variables can slow convergence. The period until the chain converges is termed the ``burn-in time.'' Once the chain has reached its steady-state, the generated samples can be treated as if they were generated from the joint distribution directly.

\subsubsection{Slice Sampling} 
\begin{figure}[!t]
\centering
\includegraphics[width=3in]{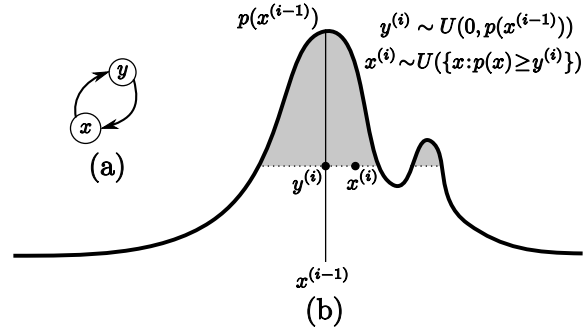}
\caption{Slice sampling of $p(x)$ illustrated: (a) Sampling is performed by traversing a Markov chain to approximate $p(x)$, the stationary distribution. Each iteration consists of (b) uniformly choosing a slice $\{x : p(x) = y\}$ and uniformly picking a new sample $x$ from that slice~\cite{WellerThesis}.}
\label{fig:slicesampling}
\end{figure}

Slice sampling is a Markov chain Monte Carlo method described in~\cite{Neal03} for sampling from $p(x)$ by sampling uniformly from the region under the pdf. Considering the area under the pdf as a pair of random variables $(X,Y)$, where $X$ is as before, and $Y \in (0,p(X))$, with joint distribution $p(x,y)$. As in Figure~\ref{fig:slicesampling}, the joint pdf can be sampled by repeatedly drawing from the conditional distributions, like in Gibbs sampling. These two conditional densities correspond exactly to uniform distributions, the first $p(y \mid x)$ being over a single interval, and the second $p(x \mid y)$ over the ``slice.''

Sampling from the slice $\{x : p(x) > y^{(i+1)}\}$ is generally difficult. If $X$ is bounded, one could simply uniformly sample over the whole interval until the generated sample has sufficiently high probability $p(x)$ to be in the slice. If the derivatives of $p(x)$ are available, root-finding methods such as Newton's method or Halley's method may be used to locate boundaries of the slice. ``Shrinkage,'' a simple accept-reject method to sample from the slice described in~\cite{Neal03}, is used in this paper.

\section{Classical (Non-Random) Parameter Estimation}\label{sec:classical}

Minimizing the MSE without access to $p(\mathbf{x})$ is difficult because the optimal estimator must effectively minimize the MSE for all possible values of $\mathbf{x}$. Moreover, the MSE of a candidate estimator generally depends on $\mathbf{x}$, so it is difficult even to evaluate the performance of any given estimator directly. As a basis of comparison, the Cram\'{e}r-Rao bound (CRB) can provide a lower bound on the MSE performance of unbiased estimators, given that we can evaluate this bound for the noisy observation model.

The best linear unbiased estimator (BLUE) is explored as a first attempt to attain the Cram\'{e}r-Rao bound. To improve upon linear estimation, the ML estimator is derived, since it is guaranteed to asymptotically achieve the CRB as the oversampling factor, $M$, goes to infinity. While the likelihood function in~\eqref{eq:likelihood} is separable, it has no closed form:
\begin{equation}
p(\mathbf{y};\mathbf{x}) = \prod_{n=0}^{N-1} \int \mathcal{N}_{y_{n}}(\mathbf{h}_{n}^{T}(z_{n})\mathbf{x},\sigma_{w}^{2})\mathcal{N}_{z_{n}}(0,\sigma_{z}^{2})\,dz_{n},\label{eq:nrlsep}
\end{equation}
where $\mathbf{h}_{n}^{T}(z_{n})$ is the $n$th row vector of the matrix $\mathbf{H}(\mathbf{z})$. An EM algorithm is developed to simplify the problem of maximizing the likelihood function.

\subsection{Cram\'{e}r-Rao Lower Bound}

The Cram\'{e}r-Rao lower bound on the MSE for the class of unbiased estimators is equal to the trace of $\mathbf{I_{y}}(\mathbf{x})^{-1}$, where the Fisher information matrix $\mathbf{I_{y}}(\mathbf{x})$ can be expressed as
\begin{equation}
\mathbf{I_{y}}(\mathbf{x}) \defeq \mathbb{E}\[\(\frac{\partial l(\mathbf{x};\mathbf{y})}{\partial \mathbf{x}}\)\(\frac{\partial l(\mathbf{x};\mathbf{y})}{\partial \mathbf{x}}\)^{T}\] = -\mathbb{E}\[\frac{\partial^{2} l(\mathbf{x};\mathbf{y})}{\partial \mathbf{x}\partial \mathbf{x}^{T}}\],\label{eq:nrfisher}
\end{equation}
and $l(\mathbf{x};\mathbf{y}) \defeq \ln p(\mathbf{y};\mathbf{x})$ is the log-likelihood of $\mathbf{y}$ as a function $\mathbf{x}$. An efficient estimator, which satisfies the CRB, can be found for the the no-jitter ($\mathbf{z} = 0$) case using the general formula in~\cite{Kay93}:
\begin{equation}
\mathbf{\hat{x}}_{\text{eff}\mid\mathbf{z} = \mathbf{0}}(\mathbf{y}) = (\mathbf{H}(\mathbf{0})^{T}\mathbf{H}(\mathbf{0}))^{-1}\mathbf{H}(\mathbf{0})^{T}\mathbf{y}.\label{eq:nrnozefficientest}
\end{equation}
However, for the random jitter case, the efficient estimator would depend on $\mathbf{x}$, so it is not valid.

For the random jitter case, the lack of a closed-form for the likelihood function requires that the lower bound be approximated numerically. Because the likelihood function is separable,
\begin{equation}
l(\mathbf{x};\mathbf{y}) = \sum_{n=0}^{N-1} \ln p(y_{n};\mathbf{x}),\label{eq:nrllsep}
\end{equation}
which means that the Fisher information in~\eqref{eq:nrfisher} can be re-written as
\begin{equation}
\mathbf{I_{y}}(\mathbf{x}) = \sum_{n=0}^{N-1} \mathbb{E}\[\(\frac{\partial\ln p(y_{n};\mathbf{x})}{\partial\mathbf{x}}\)\(\frac{\partial\ln p(y_{n};\mathbf{x})}{\partial\mathbf{x}}\)^{T}\] .\label{eq:nrfishersep}
\end{equation}
The singleton likelihood function $p(y_{n};\mathbf{x})$ can be computed numerically using Gauss-Hermite quadrature:
\begin{equation}
p(y_{n};\mathbf{x}) \approx \sum_{i=1}^{I} w_{i}\mathcal{N}_{y_{n}}(\mathbf{h}_{n}^{T}(z_{i})\mathbf{x},\sigma_{w}^{2}).\label{eq:nrlapprox}
\end{equation}
The derivative can also be approximated using quadrature:
\begin{equation}
\begin{split}
\lefteqn{\frac{\partial p(y_{n};\mathbf{x})}{\partial\mathbf{x}}}\\ &= \int {\textstyle\frac{(y_{n}-\mathbf{h}_{n}^{T}(z_{i})\mathbf{x})\mathbf{h}_{n}(z_{i})}{\sigma_{w}^{2}}}\mathcal{N}_{y_{n}}(\mathbf{h}_{n}^{T}(z_{i})\mathbf{x},\sigma_{w}^{2})\mathcal{N}_{z_{n}}(0;\sigma_{z}^{2})\,dz_{n}\\
&\approx \sum_{i=1}^{I} w_{i}{\textstyle\frac{(y_{n}-\mathbf{h}_{n}^{T}(z_{i})\mathbf{x})\mathbf{h}_{n}(z_{i})}{\sigma_{w}^{2}}}\mathcal{N}_{y_{n}}(\mathbf{h}_{n}^{T}(z_{i})\mathbf{x},\sigma_{w}^{2}).\end{split}\label{eq:nrlderivativeapprox}
\end{equation}
Since
\begin{equation}
\frac{\partial \ln p(y_{n};\mathbf{x})}{\partial\mathbf{x}} = \frac{1}{p(y_{n};\mathbf{x})}\frac{\partial p(y_{n};\mathbf{x})}{\partial\mathbf{x}},\label{eq:nrllderivative}
\end{equation}
the expression inside the expectation in~\eqref{eq:nrfishersep} becomes a complicated non-linear function of $\mathbf{y}$ and $\mathbf{x}$. Since $p(y_{n};\mathbf{x})$ is approximately a Gaussian mixture, we can resort to simple stochastic approximation to approximate this expectation, and hence, the Fisher information matrix. For each data sample $n$, generate $N_{s}$ samples $y_{s}$ from the Gaussian mixture $\sum_{i=1}^{I} w_{i}\mathcal{N}_{y}(\mathbf{h}_{n}^{T}(z_{i})\mathbf{x},\sigma_{w}^{2})$. Then,
\begin{equation}
\mathbf{I_{y}}(\mathbf{x}) \approx \sum_{n=0}^{N-1} \sum_{s=1}^{N_{s}} \mathbf{F}_{n}(y_{s};\mathbf{x}),\label{eq:nrfishersample}
\end{equation}
where
\begin{equation}
\begin{split}
\lefteqn{\mathbf{F}_{n}(y_{s};\mathbf{x})}\\ &= \(\frac{\sum_{i=1}^{I} w_{i}\mathcal{N}_{y_{s}}(\mathbf{h}_{n}^{T}(z_{i})\mathbf{x},\sigma_{w}^{2})(y_{s}-\mathbf{h}_{n}^{T}(z_{i})\mathbf{x})\mathbf{h}_{n}(z_{i})}{\sigma_{w}^{2}\sum_{i = 1}^{I} w_{i}\mathcal{N}_{y_{s}}(\mathbf{h}_{n}^{T}(z_{i})\mathbf{x},\sigma_{w}^{2})}\)\\
&\quad\cdot\(\frac{\sum_{i=1}^{I} w_{i}\mathcal{N}(y_{s};\mathbf{h}_{n}^{T}(z_{i})\mathbf{x},\sigma_{w}^{2})(y_{s}-\mathbf{h}_{n}^{T}(z_{i})\mathbf{x})\mathbf{h}_{n}(z_{i})}{\sigma_{w}^{2}\sum_{i = 1}^{I} w_{i}\mathcal{N}_{y_{s}}(\mathbf{h}_{n}^{T}(z_{i})\mathbf{x},\sigma_{w}^{2})}\)^{T}.\end{split}\label{eq:nrfishersampleFmatrix}
\end{equation}
The trace of the inverse of this matrix approximates Cram\'{e}r-Rao lower bound.

\subsection{Linear Estimation}

It is well known that for any linear observation model $\mathbf{y} = \mathbf{H}\mathbf{x}+\mathbf{w}$ with zero-mean additive noise $\mathbf{w}$ and random matrix $\mathbf{H}$, a linear estimator $\mathbf{A}\mathbf{y}$ is unbiased if and only if $\mathbf{A}\mathbb{E}[\mathbf{H}] = \mathbf{I}$. If $\mathbb{E}[\mathbf{H}]$ is full column rank, one possible linear unbiased estimator for~\eqref{eq:obsmodel} is the pseudoinverse
\begin{equation}
\mathbf{\hat{x}}_{\text{L}}(\mathbf{y}) = \mathbb{E}[\mathbf{H}(\mathbf{z})]^{\dag}\mathbf{y}.\label{eq:nrlinunbiasedest}
\end{equation}
The question naturally arises as to how to find the best linear unbiased estimator (BLUE), in terms of minimizing the MSE. In~\cite{Kay93}, the BLUE is derived for a fixed matrix $\mathbf{H}$. As derived in~\cite{WellerThesis}, the BLUE for a linear observation model with random matrix $\mathbf{H}$ is
\begin{equation}
\mathbf{\hat{x}}_{\text{BLUE}}(\mathbf{y}) = (\mathbb{E}[\mathbf{H}(\mathbf{z})]^{T}\mathbf{\Lambda_{y}}^{-1}\mathbb{E}[\mathbf{H}(\mathbf{z})])^{-1}\mathbb{E}[\mathbf{H}(\mathbf{z})]^{T}\mathbf{\Lambda_{y}}^{-1}\mathbf{y},\label{eq:nrbluest}
\end{equation}
where the covariance matrix of the data $\mathbf{\Lambda_{y}}$ depends on the value of the parameters:
\begin{equation}
\mathbf{\Lambda_{y}} = \mathbb{E}[\mathbf{H}(\mathbf{z})\mathbf{x}\mathbf{x}^{T}\mathbf{H}(\mathbf{z})^{T}] - \mathbb{E}[\mathbf{H}(\mathbf{z})]\mathbf{x}\mathbf{x}^{T}\mathbb{E}[\mathbf{H}(\mathbf{z})]^{T}+\sigma_{w}^{2}\mathbf{I}.\label{eq:nrycov}
\end{equation}
A sufficient condition for~\eqref{eq:nrbluest} to be valid is for $\mathbf{\Lambda_{y}}$ to be a scalar matrix (a scalar times the identity matrix). In general, and for the observation model and prior on the jitter used here, the diagonal elements of $\mathbf{\Lambda_{y}}$ are not equal, and $\mathbf{\Lambda_{y}}$ is not a scalar matrix. For the no-jitter case, $\mathbf{H}(\mathbf{z})$ is deterministic, and the BLUE is equal to the efficient linear estimator in~\eqref{eq:nrnozefficientest}.

This efficient linear estimator for the no-jitter case is used as a baseline to compare the performance of the ML estimator proposed in the next section. This estimator would be optimal in the MSE sense if the jitter were removed from the observation model.

\subsection{ML Estimation using the EM Algorithm}

Operating under the assumption that a linear estimator is too restrictive to capture the complicated behavior of the random jitter observation model, but unable to formulate an efficient estimator directly, maximum likelihood estimation appears to be a promising alternative. However, maximizing the likelihood function is a difficult task, because the likelihood function is not convex in $\mathbf{x}$, does not have a closed form, and is not separable. One approach would be to utilize a gradient method, or a hill-climbing method, to find a local maximum iterating over one variable at a time. One related method, investigated in~\cite{Kusuma08}, involves maximizing the joint likelihood $p(\mathbf{y},\mathbf{z};\mathbf{x})$ by alternating maximizing $p(\mathbf{z}\mid \mathbf{y};\mathbf{x})$ and $p(\mathbf{y}\mid \mathbf{z};\mathbf{x})$.

Maximum likelihood estimation is also a classic application of the Expectation-Maximization (EM) algorithm, as described in~\cite{Dempster77}. For each iteration $i$, the EM algorithm consists of maximizing the expectation
\begin{equation}
Q(\mathbf{x};\mathbf{\hat{x}}^{(i-1)}) = \mathbb{E}\[\log p(\mathbf{y},\mathbf{z};\mathbf{x})\mid \mathbf{y};\mathbf{\hat{x}}^{(i-1)}\]\label{eq:nremexpect}
\end{equation}
with respect to the unknown parameters $\mathbf{x}$; i.e. $\mathbf{\hat{x}}^{(i)} = \argmax_{\mathbf{x}} Q(\mathbf{x};\mathbf{\hat{x}}^{(i-1)})$. Here, $\mathbf{z}$ is the missing data. Repeated iterations of this step results in the maximizing $\mathbf{x}$ converging to a local maximum of the likelihood function.

Maximizing~\eqref{eq:nremexpect} will result in finding a local maximum of the likelihood function; however, nothing has been said about how fast the algorithm converges to a stationary point, or if this local maximum coincides with the true ML estimate. The rate of convergence depends on the choice of complete data~\cite{Dempster77}. Proven in~\cite{Herzet07} is the fact that the rate of convergence is worse if the CRB for the incomplete data set is much greater than the CRB for the augmented data set. The SEM algorithm in~\cite{Meng91} also approximates the observed Fisher information, which can be used to establish the quality of the estimate. Since the likelihood function is not concave in $\mathbf{x}$, the estimate depends on initial conditions, the effects of which are studied in~\cite{WellerThesis}.

If the values of the jitter were known, the problem would decompose into a linear one, and the solution is readily available. Then, selecting the jitter to be the missing data,
\begin{equation}
\ln p(\mathbf{y},\mathbf{z};\mathbf{x}) = -\frac{1}{2\sigma_{w}^{2}}\|\mathbf{y}-\mathbf{H}(\mathbf{z})\mathbf{x}\|_{2}^{2} - \frac{1}{2\sigma_{z}^{2}}\|\mathbf{z}\|_{2}^{2} - N\log (2\pi\sigma_{w}\sigma_{z}).\label{eq:nremjointlikelihood}
\end{equation}
Expanding and substituting into the expectation in~\eqref{eq:nremexpect} yields
\begin{equation}
\begin{split}
Q(\mathbf{x};\mathbf{\hat{x}}^{(i-1)}) &= \frac{-1}{2\sigma_{w}^{2}}\(\mathbf{y}^{T}\mathbf{y}-2\mathbf{y}^{T}\mathbb{E}\[\mathbf{H}(\mathbf{z})\mid \mathbf{y};\mathbf{\hat{x}}^{(i-1)}\]\mathbf{x}\right.\\
&\quad \left.+ \mathbf{x}^{T}\mathbb{E}\[\mathbf{H}(\mathbf{z})^{T}\mathbf{H}(\mathbf{z})\mid \mathbf{y};\mathbf{\hat{x}}^{(i-1)}\]\mathbf{x}\)\\
&\quad- \frac{1}{2\sigma_{z}^{2}}\mathbb{E}\[\mathbf{z}^{T}\mathbf{z}\mid \mathbf{y};\mathbf{\hat{x}}^{(i-1)}\]  - N\log (2\pi\sigma_{w}\sigma_{z}).\end{split}\label{eq:nremQ}
\end{equation}
Since this equation is quadratic in $\mathbf{x}$, the maximum value $\mathbf{\hat{x}}^{(i)}$ satisfies the linear system of equations
\begin{equation}
\mathbb{E}\[\mathbf{H}(\mathbf{z})^{T}\mathbf{H}(\mathbf{z})\mid \mathbf{y};\mathbf{\hat{x}}^{(i-1)}\]\mathbf{x}^{(i)} = \mathbb{E}\[\mathbf{H}(\mathbf{z})\mid \mathbf{y};\mathbf{\hat{x}}^{(i-1)}\]^{T}\mathbf{y}.\label{eq:nremxsolution}
\end{equation}
The Hessian matrix is negative-definite, so the equation is strictly concave, and this maximum is unique. To solve this linear system of equations, the expectations in~\eqref{eq:nremxsolution} need to be approximated. Using Bayes rule,
\begin{equation}
p(\mathbf{z}\mid \mathbf{y};\mathbf{\hat{x}}^{(i-1)}) = \prod_{n=0}^{N-1}\frac{p(y_{n}\mid z_{n};\mathbf{\hat{x}}^{(i-1)})p(z_{n})}{p(y_{n};\mathbf{\hat{x}}^{(i-1)})}.\label{eq:nrposteriorbayes}
\end{equation}
If the jitter or additive noise were not independent (white), the expectation would not be separable. Using Gauss-Hermite quadrature (with $I$ terms) on the left and right expectations,
\begin{align}
\begin{split}
\lefteqn{\mathbb{E}\[\mathbf{H}(\mathbf{z})^{T}\mathbf{H}(\mathbf{z})\mid \mathbf{y};\mathbf{\hat{x}}^{(i-1)}\]}\\ &= \sum_{n=0}^{N-1}\mathbb{E}\[\mathbf{h}_{n}(z_{n})\mathbf{h}_{n}^{T}(z_{n})\mid y_{n};\mathbf{\hat{x}}^{(i-1)}\]\\
&\approx \sum_{n=0}^{N-1} \frac{1}{p(y_{n};\mathbf{\hat{x}}^{(i-1)})}\sum_{j=1}^{I} w_{j}\mathbf{h}_{n}(z_{j})\mathbf{h}_{n}^{T}(z_{j})p(y_{n}\mid z_{j};\mathbf{\hat{x}}^{(i-1)}),\end{split}\label{eq:nremxsolnleftquad}\displaybreak[1]\\
\begin{split}
\lefteqn{\[\mathbb{E}\[\mathbf{H}(\mathbf{z})\mid \mathbf{y};\mathbf{\hat{x}}^{(i-1)}\]\]_{n,:}}\\ &= \mathbb{E}\[\mathbf{h}_{n}^{T}(z_{n})\mid y_{n};\mathbf{\hat{x}}^{(i-1)}\]\\
&\approx \frac{1}{p(y_{n};\mathbf{\hat{x}}^{(i-1)})}\sum_{j=1}^{I} w_{j}\mathbf{h}_{n}^{T}(z_{j})p(y_{n}\mid z_{j};\mathbf{\hat{x}}^{(i-1)}),\end{split}\label{eq:nremxsolnrightquad}
\end{align}
and $p(y_{n};\mathbf{x}^{(i-1)})$ is computed again using~\eqref{eq:nrlapprox}. The complexity of performing these approximations is linear in the number of samples, so for a large number of parameters, the limiting step is actually solving the system once the expectations are computed.

\section{Bayesian Estimation}\label{sec:bayesian}

The classical estimation framework is suitable for designing a general post-processor for a sampling system when nothing is known about the process generating the parameters of the input signal. However, if the generating process is known, Bayesian inference leverages that information to optimize the performance of the post-processor. The Bayes least squares (BLS) estimator minimizes the MSE, and approximating that estimator is the objective of this section.

How should the prior distribution be chosen? A white Gaussian process is attractive for its simplicity. The conjugate prior is another prior employed for convenience. However, since the likelihood function is already intractable, a conjugate prior is unhelpful for this problem. The ``least-informative'' prior is another suitable choice, but in general, it is hard to find. The approach used here is the maximum entropy prior. The maximum entropy model is justified in that it imposes minimal structure on the parameters; in this way, the maximum entropy model is the most ``challenging'' prior to use. Therefore, the parameters should be independent, and the entropy-maximizing distribution over a finite interval is the uniform prior~\cite{Cover06}. For simplicity, we will assume that the parameters lie between $-1$ and $1$.

Like the likelihood function in the previous section, the resulting posterior density has no closed form. It can be expressed in terms of the singleton likelihoods in~\eqref{eq:nrlsep}:
\begin{equation}
p(\mathbf{x}\mid \mathbf{y}) = \frac{\prod_{n=0}^{N-1} p(y_{n}\mid \mathbf{x})\prod_{k=0}^{K-1}p(x_{k})}{\int\cdots\int \prod_{n=0}^{N-1} p(y_{n}\mid \mathbf{x'})\prod_{k=0}^{K-1}p(x_{k}')\,d\mathbf{x'}}.\label{eq:bayesposteriorsep}
\end{equation}

\subsection{LLS Estimation}

The linear least squares (LLS) estimator is defined to be the linear estimator with minimum MSE. For the random jitter observation model in~\eqref{eq:obsmodel}, the LLS estimator is
\begin{equation}
\mathbf{\hat{x}}_{\text{LLS}}(\mathbf{y}) = \mathbb{E}[\mathbf{H}(\mathbf{z})]^{T}\(\mathbb{E}[\mathbf{H}(\mathbf{z})\mathbf{H}(\mathbf{z})^{T}]+\frac{\sigma_{w}^{2}}{\sigma_{x}^{2}}\mathbf{I}\)^{-1}\mathbf{y}.\label{eq:bayesllsest}
\end{equation}
The expectations in~\eqref{eq:bayesllsest} can be computed using Gauss-Hermite quadrature or some other numerical method. However, the optimal linear estimator has already been studied and does not show much improvement over the LLS estimator assuming no jitter is present. In~\cite{Balakrishnan62}, the optimal filter coefficients for a low pass interpolator are derived for the random jitter problem. As mentioned in the introduction, the MSE of a low pass interpolator is linear in the variance of the jitter~\cite{Liu65}.

When no jitter is assumed, the above estimator simplifies to
\begin{equation}
\mathbf{\hat{x}}_{\text{LLS}\mid\mathbf{z}=\mathbf{0}}(\mathbf{y}) = \mathbf{H}(\mathbf{0})^{T}\(\mathbf{H}(\mathbf{0})\mathbf{H}(\mathbf{0})^{T}+\frac{\sigma_{w}^{2}}{\sigma_{x}^{2}}\mathbf{I}\)^{-1}\mathbf{y}.\label{eq:bayesllsestnoz}
\end{equation}
This linear estimator is the best linear operation that can be performed in the absence of jitter. Hence, the no-jitter LLS estimator is the baseline estimator against which the BLS estimators derived later will be measured. The error covariance of this estimator is
\begin{equation}
\begin{split}
\mathbf{\Lambda}_{\text{LLS}\mid\mathbf{z}=\mathbf{0}} &= \sigma_{x}^{2}\(\mathbf{I} - \mathbf{H}(\mathbf{0})^{T}\(\mathbf{H}(\mathbf{0})\mathbf{H}(\mathbf{0})^{T}+\frac{\sigma_{w}^{2}}{\sigma_{x}^{2}}\mathbf{I}\)^{-1}\mathbf{H}(\mathbf{0})\)\\
&= \frac{\sigma_{x}^{2}}{1 + M\sigma_{x}^{2}/\sigma_{w}^{2}}\mathbf{I} = \frac{\sigma_{w}^{2}}{M + \sigma_{w}^{2}/\sigma_{x}^{2}}\mathbf{I}.\end{split}\label{eq:bayesllsnozerrcov}
\end{equation}

\subsection{BLS Estimation using Gibbs Sampling}

The BLS estimator is well-known to have the minimum MSE of any estimator for the Bayesian framework. However, optimal performance comes at a price: the BLS estimator involves taking the expected value of the posterior distribution, which is generally difficult to compute, as it is for this problem. However, stochastic approximation can be helpful here. Consider the BLS estimator for $(\mathbf{x},\mathbf{z})$. The expectation $\mathbb{E}[\mathbf{x},\mathbf{z}\mid \mathbf{y}]$ is an obvious application of Gibbs sampling. Sampling from $p(\mathbf{x},\mathbf{z}\mid \mathbf{y})$ directly is a terrible idea due to the high dimensionality of the sample space, even if it were trivial to generate samples. Instead, Gibbs sampling can be used to generate samples of one variable at a time, conditioned on the rest.

Once enough samples have been taken so that the Markov chain is sufficiently close to the stationary distribution, additional samples are averaged to approximate the BLS estimator. In the description below, $I_{b}$ represents the ``burn-in time'', the number of iterations until the Markov chain has reached its steady state, and $I$ represents the number of samples to generate after convergence has been achieved.
\begin{small}
\begin{algorithmic}
\REQUIRE $\mathbf{y},I,I_{b}$
\STATE $\mathbf{z}^{(0)} = \mathbf{0}$, $\mathbf{x}^{(0)} = \mathbf{0}$
\FOR{$i = 1:I+I_{b}$}
\STATE \begin{tabular}{rcl}
$z_{0}^{(i)}$ &$\sim$& $p(\cdot\mid z_{1}^{(i-1)},\ldots,z_{N-1}^{(i-1)},\mathbf{x}^{(i-1)},\mathbf{y})$\\
$z_{1}^{(i)}$ &$\sim$& $p(\cdot\mid z_{0}^{(i)},z_{2}^{(i-1)},\ldots,z_{N-1}^{(i-1)},\mathbf{x}^{(i-1)},\mathbf{y})$\\
&\vdots&\\
$z_{N-1}^{(i)}$ &$\sim$& $p(\cdot\mid z_{0}^{(i)},\ldots,z_{N-2}^{(i)},\mathbf{x}^{(i-1)},\mathbf{y})$\\
$x_{0}^{(i)}$ &$\sim$& $p(\cdot\mid \mathbf{z}^{(i)},x_{1}^{(i-1)},\ldots,x_{K-1}^{(i-1)},\mathbf{y})$\\
$x_{1}^{(i)}$ &$\sim$& $p(\cdot\mid \mathbf{z}^{(i)},x_{0}^{(i)},x_{2}^{(i-1)},\ldots,x_{K-1}^{(i-1)},\mathbf{y})$\\
&\vdots&\\
$x_{K-1}^{(i)}$ &$\sim$& $p(\cdot\mid \mathbf{z}^{(i)},x_{0}^{(i)},\ldots,x_{K-2}^{(i)},\mathbf{y})$
\end{tabular}
\ENDFOR
\STATE $\mathbf{\hat{x}} = \frac{1}{I}\sum_{i=I_{b}+1}^{I_{b}+I} \mathbf{x}^{(i)}$, $\mathbf{\hat{z}} = \frac{1}{I}\sum_{i=I_{b}+1}^{I_{b}+I} \mathbf{z}^{(i)}$
\RETURN $\mathbf{\hat{x}}$, $\mathbf{\hat{z}}$
\end{algorithmic}
\end{small}

Consider generating samples $z_{n}$ from the distribution $p(\cdot\mid \mathbf{z}_{-n},\mathbf{x},\mathbf{y})$. Using Bayes rule, and the independence of $z_{n}$ and $w_{n}$,
\begin{equation}
\begin{split}
p(z_{n}\mid \mathbf{z}_{-n},\mathbf{x},\mathbf{y}) &= \frac{p(\mathbf{y}\mid \mathbf{z},\mathbf{x})p(\mathbf{z})p(\mathbf{x})}{p(\mathbf{z}_{-n},\mathbf{x},\mathbf{y})}\\
&\propto \mathcal{N}_{y_{n}}(\mathbf{h}_{n}^{T}(z_{n})\mathbf{x},\sigma_{w}^{2})\mathcal{N}_{z_{n}}(0,\sigma_{z}^{2}).\end{split}\label{eq:bayesznpdf}
\end{equation}
The full-conditional distribution of $z_{n}$ is independent of $\mathbf{z}_{-n}$, so sampling the jitter values can be easily grouped together. Because this functional form is enveloped by the prior on $z_{n}$, rejection sampling is an obvious choice to produce samples. The proposal density is $q(z_{n}) = \mathcal{N}_{z_{n}}(0,\sigma_{z}^{2})$, and the scaling factor is $c = 1/\sqrt{2\pi\sigma_{w}^{2}}$. In fact, since $\mathbf{h}_{n}^{T}(z_{n})\mathbf{x} \leq \|\mathbf{h}_{n}(z_{n})\|_{2}\|\mathbf{x}\|_{2} = (1)\|\mathbf{x}\|_{2}$,
\begin{equation}
\begin{split}
\lefteqn{\mathcal{N}_{y_{n}}(\mathbf{h}_{n}^{T}(z_{n})\mathbf{x},\sigma_{w}^{2})}\\ &= \frac{1}{\sqrt{2\pi\sigma_{w}^{2}}}\exp\[-\frac{y_{n}^{2} - 2y_{n}\mathbf{h}_{n}^{T}(z_{n})\mathbf{x} + (\mathbf{h}_{n}^{T}(z_{n})\mathbf{x})^{2}}{2\sigma_{w}^{2}}\]\\
&\leq \frac{1}{\sqrt{2\pi\sigma_{w}^{2}}}\exp\[-\frac{y_{n}^{2} - 2y_{n}\|\mathbf{x}\|_{2}}{2\sigma_{w}^{2}}\],\end{split}\label{eq:bayeszncalt}
\end{equation}
and if $y_{n}^{2} > 2y_{n}\|\mathbf{x}\|_{2}$, $c$ can be multiplied by $\exp\[-\frac{y_{n}^{2} - 2y_{n}\|\mathbf{x}\|_{2}}{2\sigma_{w}^{2}}\]$, and $cq(z_{n})$ would more tightly envelope~\eqref{eq:bayesznpdf}.

Unlike the independent $z_{n}$'s, the conditional distribution on $x_{k}$ does depend on the other parameters $\mathbf{x}_{-k}$:
\begin{equation}
\begin{split}
p(x_{k}\mid \mathbf{x}_{-k},\mathbf{z},\mathbf{y}) &= \frac{p(\mathbf{y}\mid \mathbf{z},\mathbf{x})p(\mathbf{z})p(\mathbf{x})}{p(\mathbf{x}_{-k},\mathbf{z},\mathbf{y})}\\
&\propto \mathcal{N}_{\mathbf{y}}(\mathbf{H}(\mathbf{z})\mathbf{x},\sigma_{w}^{2}\mathbf{I})U_{x_{k}}(-1,1).\end{split}\label{eq:bayesxkpdf}
\end{equation}
Denote by $\mathbf{H}_{k}(\mathbf{z})$ and $\mathbf{H}_{-k}(\mathbf{z})$ the columns of $\mathbf{H}(\mathbf{z})$ that multiply $x_{k}$ and $\mathbf{x}_{-k}$ in~\eqref{eq:bayesxkpdf}, respectively. The truncated normal distribution in~\eqref{eq:bayesxkpdf} can be modified to be a distribution explicitly on $x_{k}$:
\begin{equation}
x_{k} \sim \mathcal{N}_{x_{k}}(\mu_{k},\sigma^{2}_{k})U_{x_{k}}(-1,1),\label{eq:bayesxkpdf2}
\end{equation}
where
\begin{equation}
\mu_{k} = \frac{\mathbf{H}_{k}(\mathbf{z})^{T}(\mathbf{y} - \mathbf{H}_{-k}(\mathbf{z})\mathbf{x}_{-k})}{\mathbf{H}_{k}(\mathbf{z})^{T}\mathbf{H}_{k}(\mathbf{z})}\label{eq:bayesgibbsmusigmaxk},\quad\sigma^{2}_{k} = \frac{\sigma_{w}^{2}}{\mathbf{H}_{k}(\mathbf{z})^{T}\mathbf{H}_{k}(\mathbf{z})}.
\end{equation}

Sampling from a truncated normal distribution is a typical application of cdf inversion. When the mean $\mu_{k}$ lies between $-1$ and $1$, the cdf is almost linear, and the cdf inversion technique works well. However, when the mean lies far outside this range, the cdf is practically flat, and the technique is limited by the precision of the computer. A rejection sampling-based approach for producing samples of a truncated standard normal distribution is developed in~\cite{Robert95}. For a standard normal distribution truncated to the interval $[a,b]$, where $a > 0$, the proposal distribution should be an exponential distribution shifted to cover $x \geq a$, whose optimal scale factor
\begin{equation}
\alpha^{*} = \frac{a + \sqrt{a^{2} + 4}}{2}.\label{eq:alphaopt}
\end{equation}
The case where $[a,b]$ lies to the left of zero, generating samples is the mirror-image of this problem, so the optimal exponential distribution is flipped and shifted to cover $x \leq b$, and $\alpha^{*} = -\frac{1}{2}(b-\sqrt{b^{2}+4})$. According to~\cite{Robert95}, a uniform distribution $U_{x}(a,b)$ is the preferred proposal distribution when $a$ and $b$ are too close together, so that a disproportionately large number of samples are not rejected due to the $x \leq b$ or $x \geq a$ requirement.

The parameters $x_{k}$ are all highly correlated, so one might wonder if it is possible to use a similar technique to generate realizations of the multivariate truncated normal distribution $p(\mathbf{x}\mid \mathbf{y},\mathbf{z})$ directly. However, as~\cite{Robert95} laments, the rejection-sampling method has no natural extension to multivariate truncated normal distributions when the variables are correlated. Gibbs sampling is recommended to reduce the problem down to univariate sampling, which is what is proposed here anyway. Sampling $x_{k}$ one at a time has the added benefit of extending naturally for a less trivial generating process for $\mathbf{x}$.

\subsection{Improvement via Slice Sampling}

\begin{figure}[!t]
\centering
\includegraphics[width=3in]{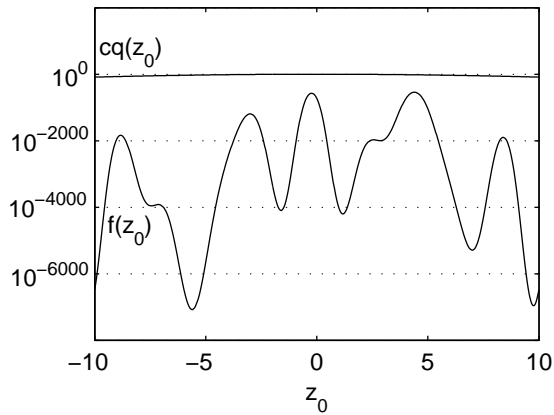}
\caption{Example plot of the unnormalized pdf $f(z_{0}) = \tilde{p}(z_{0}\mid \mathbf{y},\mathbf{x})$, proportional to $p(z_{0}\mid \mathbf{x},\mathbf{y})$, with respect to $z_{0}$, for $K = 10$, $\sigma_{w} = 0.01$, $\sigma_{z} = 0.5$, and $M = 16$. The proposal distribution for rejection sampling (a zero-mean normal distribution with variance $\sigma_{z}^{2}$) is also shown for reference. Note that even for values of $z_{0}$ likely according to the proposal density $q(z_{0})$, the target density is many orders of magnitude smaller, so rejection sampling will reject an extraordinarily large number of samples before accepting a sample.}
\label{fig:pzngivenynx}
\end{figure}

While the rejection sampler provides an exact method for producing realizations of a target distribution, a significant disparity between the shape of the proposal and target distributions causes a large number of samples to be rejected. In the case of rejection sampling employed in the Gibbs sampler to generate jitter values $z_{n}$ from the density~\eqref{eq:bayesznpdf}, finding the smallest value for $c$ is difficult, and the shape of the target distribution can vary depending on the parameters. Empirical evidence, shown in Figure~\ref{fig:pzngivenynx}, portrays the extent of the problem with rejection sampling, when $\sigma_{z} \gg \sigma_{w}$. High oversampling only compounds this phenomenon.

Slice sampling is not susceptible to such problems since no tightly enveloping proposal density is necessary; the ability to evaluate an unnormalized form of the target distribution is sufficient. Thus, the expression in~\eqref{eq:bayesznpdf} can still be used. Each iteration of slice sampling consists of two uniform sampling problems:
\begin{enumerate}
\item Choose a slice $u$ uniformly from $[0,\tilde{p}(z_{n}^{(i)}\mid \mathbf{y},\mathbf{x})]$.
\item Sample $z_{n}^{(i+1)}$ uniformly from the slice $S \defeq \{z_{n} : \tilde{p}(z_{n}\mid \mathbf{y},\mathbf{x}) \geq u\}$.
\end{enumerate}
The first step is trivial, since we are sampling from a single interval. The second step is more difficult. However, since $u \leq \tilde{p}(z_{n}\mid \mathbf{y},\mathbf{x})$ for all $z_{n}$ in the slice,
\begin{equation}
\begin{split}
\log u &\leq -\frac{(y_{n} - \mathbf{h}_{n}^{T}(z_{n})\mathbf{x})^{2}}{2\sigma_{w}^{2}} - \frac{z_{n}^{2}}{2\sigma_{z}^{2}} - \log (2\pi\sigma_{z}\sigma_{w})\\
&\leq - \frac{z_{n}^{2}}{2\sigma_{z}^{2}} - \log (2\pi\sigma_{z}\sigma_{w}).\end{split}\label{eq:bayessliceconstraint}
\end{equation}
Solving for $z_{n}$, the range of possible $z_{n}$ is bounded:
\begin{equation}
|z_{n}| \leq \sigma_{z}\sqrt{-2\log u - 2\log(2\pi\sigma_{w}\sigma_{z})}.\label{eq:bayessliceznrange}
\end{equation}
Using these extreme points for the initial interval containing the slice, and the ``shrinkage'' method specified in~\cite{Neal03} to sample from the slice by repeatedly shrinking the interval, slice sampling becomes a relatively efficient alternative to rejection sampling. The ``shrinkage'' method decreases the size of the interval exponentially fast, on average. Consider one iteration of shrinkage, where the previous point $x_{0}$ lies in the interval $[L,R]$. Then, the expected size of the new interval $[L',R']$ is
\begin{equation}
\begin{split}
\mathbb{E}[R'-L'\mid R,L,x_{0}] &= \frac{1}{R-L}\[\int_{L}^{x_{0}} (R - x)\,dx + \int_{x_{0}}^{R} (x-L)\,dx\]\\
&= \frac{R^{2}-2RL+L^{2}}{2(R-L)} + \frac{x_{0}(R+L-x_{0}) - RL}{R-L}\\
&= \frac{R-L}{2} + \frac{x_{0}(R+L-x_{0}) - RL}{R-L}.\end{split}\label{eq:bayessliceshrinkival1}
\end{equation}
This expectation is quadratic in $x_{0}$, so the maximum occurs at the extreme point $x_{0} = (R+L)/2$. The maximum value is
\begin{equation}
\begin{split}
\lefteqn{\max_{x_{0}} \mathbb{E}[R'-L'\mid R,L,x_{0}]}\\ &= \frac{R-L}{2} + \frac{((R+L)/2)(R+L-(R+L)/2) - RL}{R-L}\\
&= \frac{R-L}{2} + \frac{(R+L)^{2}/4 - RL}{R-L} = \frac{3}{4}(R-L).\end{split}\label{eq:bayessliceshrinkivalmax}
\end{equation}
Concavity implies that the minima are at the two endpoints $x_{0} = L$ and $x_{0} = R$. In both cases, the expected size of the interval is $(R-L)/2$. Therefore,
\begin{equation}
\frac{1}{2}(R-L) \leq \mathbb{E}[R'-L'\mid R,L,x_{0}] \leq \frac{3}{4}(R-L),\label{eq:bayessliceshrinkivalbounds}
\end{equation}
which implies that at worst, the size of the interval shrinks to $3/4$ its previous size per iteration, on average. Then, given the initial interval $[L_{0},R_{0}]$ and previous point $x_{0}$, the expected size of the interval $[L_{I},R_{I}]$ after $I$ iterations of the shrinkage algorithm is
\begin{equation}
\begin{split}
\lefteqn{\mathbb{E}[R_{I}-L_{I}\mid R_{0},L_{0},x_{0}]}\\ &= \mathbb{E}[\mathbb{E}[R_{I}-L_{I}\mid R_{0},L_{0},\ldots,R_{I-1},L_{I-1},x_{0}] \mid R_{0}, L_{0}, x_{0}]\\
&\leq \(\frac{3}{4}\)^{I}(R_{0}-L_{0}).\end{split}\label{eq:bayessliceshrinkivals}
\end{equation}
If the target distribution $p(x)$ is continuous, the algorithm is guaranteed to terminate once the search interval is small enough. Since the interval size shrinks exponentially fast, on average, the number of ``shrinkage'' iterations is approximately proportional to the log of the fraction of the initial interval contained in the slice. In a discussion included in~\cite{Neal03}, a binary search-like shrinkage algorithm is proposed that can converge faster on the slice than the algorithm used here. Incorporating such an approach to accelerate the slice sampler merits future investigation.

Using slice sampling to generate the jitter values when $\sigma_{z}$ is large relative to $\sigma_{w}$ improves the speed of the Gibbs sampler. However, the addition of new auxiliary variables through slice sampling can be expected to slow the Gibbs sampler's overall rate of convergence. Thus, both algorithms are included for simulation.

\section{Simulation Results}\label{sec:results}

The objectives of this section are: (a) to determine when jitter significantly affects the samples of a signal and (b) to demonstrate that nonlinear post-processing can improve upon linear estimation of the signal parameters. The convergence behavior of all the included algorithms, as well as their sensitivity to initial conditions, is presented in~\cite{WellerThesis}. This article focuses on the performance analysis of these algorithms, using Monte Carlo simulation to determine the MSE of each algorithm for the different parameter choices (specified in each figure). As in the rest of the paper, $\sigma_{z}$ is relative to the critical sampling period $T=1$, and $\sigma_{w}$ is relative to the scale of the signal parameters $x_{k} \sim U_{x}(-1,1)$.

\subsection{Evaluating the EM Algorithm}

\subsubsection{Convergence} The convergence properties of the EM algorithm were discussed briefly in the background section. Intuitively, increasing $M$, increasing $\sigma_{z}$, or decreasing $\sigma_{w}$ increases the additional information from knowing the missing data $\mathbf{z}$, slowing the rate of convergence. This behavior is observed in the first simulations in~\cite{WellerThesis}. In addition, these simulations suggest that the proposed algorithm converges exponentially fast.

The other major sticking point for using an EM algorithm is the local nature of the solution. If the likelihood function has many local maxima, the EM algorithm may not necessarily converge to the global maximum. Instead, the local maximum is determined by the choice of initial conditions; i.e. the initial value of the parameters $\mathbf{x}$. Simulations in~\cite{WellerThesis} suggest the sensitivity to initial conditions increases with the non-concavity of the likelihood function. Larger $M$, larger $\sigma_{z}$, or smaller $\sigma_{w}$ increase the rate and severity of oscillations in the likelihood function, precipitating more local maxima. In cases when the sensitivity is an issue, using multiple random starting points or approaches like the deterministic annealing EM algorithm~\cite{Ueda98} can help alleviate the problem.

\subsubsection{MSE Performance} The EM algorithm is compared against two linear estimators. First, to demonstrate the MSE improvement attainable by not restricting ourselves to linear estimators, the EM algorithm is pitted against the linear unbiased estimator in~\eqref{eq:nrlinunbiasedest}. However, a major motivating factor for developing these algorithms is to reduce the power consumption due to clock accuracy. By comparing the EM algorithm against the optimal linear estimator for the no-jitter observation model, the EM algorithm can be shown to achieve the same MSE as the linear estimator for a substantially larger jitter variance, reducing the clock's power consumption.

\begin{figure}[!t]
\centering
\includegraphics[width=3in]{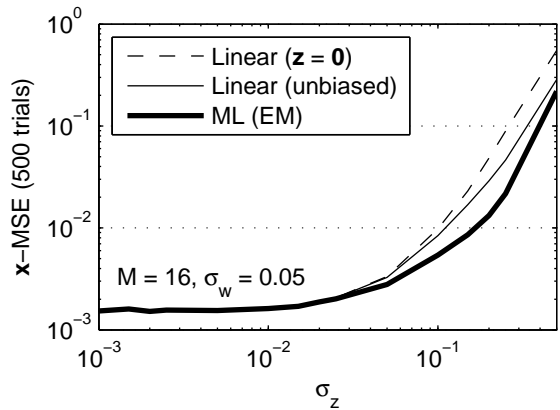}
\caption{EM Algorithm Performance: This plot compares the EM algorithm to the linear unbiased estimator in~\eqref{eq:nrlinunbiasedest} and the no-jitter efficient linear estimator in~\eqref{eq:nrnozefficientest} for $K = 10$, $M = 16$ and $\sigma_{w} = 0.05$. More extensive plots can be found in~\cite{WellerThesis}.}
\label{fig:emcomplin}
\end{figure}

When the additive noise dominates the jitter ($\sigma_{z} \ll \sigma_{w}$), the improvement can be expected to be minimal, since the system is nearly linear, and the jitter is statistically insignificant. As the amount of jitter increases, the density function $p(\mathbf{z}\mid \mathbf{y};\mathbf{x})$ used in each iteration of the EM algorithm becomes more non-linear in $\mathbf{z}$, and the quadrature becomes less accurate for a given number of terms. Therefore, the EM algorithm generally takes longer to converge, and the result should be a less accurate approximation to the true ML estimator. This behavior can be observed in Figure~\ref{fig:emcomplin}, where the EM algorithm is compared against both the linear unbiased estimator in~\eqref{eq:nrlinunbiasedest} and the efficient no-jitter linear estimator in~\eqref{eq:nrnozefficientest}. The linear unbiased estimator has lower MSE for higher $M$, and the EM algorithm generally has lower MSE than either linear estimator.

\begin{figure}[!t]
\centering
\includegraphics[width=3in]{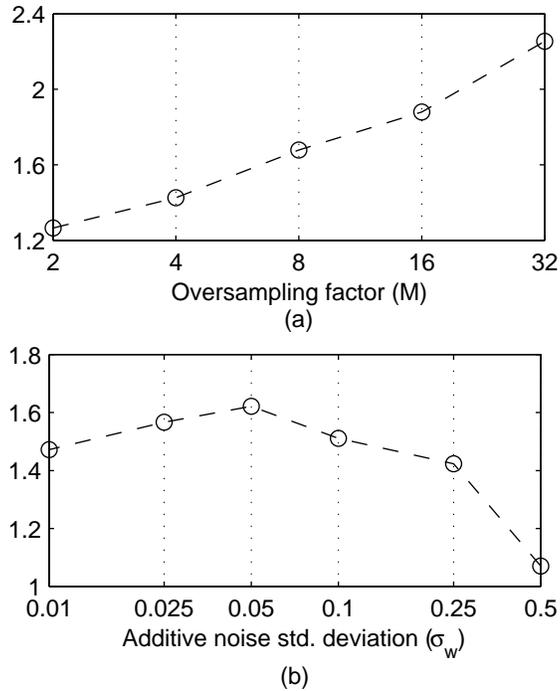}
\caption{EM Algorithm Performance: These graphs show the maximum factor of improvement in jitter tolerance, measured by $\sigma_{z}$, achievable by the EM algorithm. Holding $\sigma_{w} = 0.1$ fixed, (a) shows the trend in maximum improvement as $M$ increases, and (b) shows the trend in maximum improvement as $\sigma_{w}$ increases holding $M = 8$ fixed.}
\label{fig:emsigmazimprovement}
\end{figure}

To answer the question of how much more jitter can be tolerated for the same desired MSE using the EM algorithm, the maximum proportional increase is plotted as a function of $M$ and $\sigma_{w}$ in Figure~\ref{fig:emsigmazimprovement}. The maximum proportional increase for a choice of $M$ and $\sigma_{w}$ is computed by approximating log-log domain MSE curves, like those in Figure~\ref{fig:emcomplin}, with piece-wise linear curves and interpolating the maximum distance between them over the range of $\sigma_{z}$ up to $0.5$. The proportion of improvement increases logarithmically as $M$ increases, and the improvement stays approximately the same for different values of $\sigma_{w}$. However, the maximum improvement shown for $\sigma_{w} = 0.5$ is low because no $\sigma_{z} > 0.5$ is tested. The maximum improvement factor shown corresponds to the power consumption dropping by $50$ to $75$ percent.

\begin{figure}[!t]
\centering
\includegraphics[width=3in]{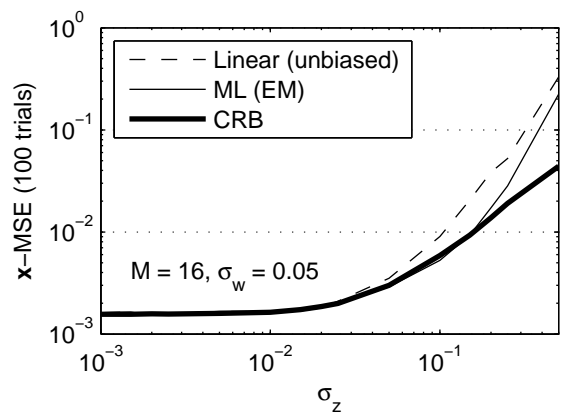}
\caption{EM Algorithm Performance: For a randomly chosen $\mathbf{x}$ ($K = 10$), the Cram\'{e}r-Rao lower bound is plotted for various choices of $\sigma_{z}$, and $M = 16$ and $\sigma_{w} = 0.05$. Then, the MSE of 100 trials of both the linear unbiased estimator and the EM algorithm are plotted against the CRB. Similar curves for $M = 2$ and $M = 4$ can be found in~\cite{WellerThesis}.}
\label{fig:emlincrbcomp}
\end{figure}

As a final experiment, the Cram\'{e}r-Rao lower bound is compared to the unbiased linear estimator~\eqref{eq:nrlinunbiasedest} and EM algorithm to measure the efficiency of the algorithms. Although computational difficulties prevent a complete comparison for every possible value of $\mathbf{x}$, carrying out a comparison for a few randomly chosen values of $\mathbf{x}$ provide a measure of the quality of the algorithms. As the curves in Figure~\ref{fig:emlincrbcomp} demonstrate for one such random choice of $\mathbf{x}$, both algorithms are approximately efficient for small $\sigma_{z}$, but the EM algorithm continues to be efficient for larger values of $\sigma_{z}$ than the linear estimator. Since the ML estimator is asymptotically efficient, as $M \rightarrow \infty$, the EM algorithm should be efficient for all $\sigma_{z}$ with large enough oversampling.

\subsection{The Gibbs Samplers}

\subsubsection{Convergence} The Gibbs sampler is known to converge if the Markov chain is irreducible and aperiodic; however, the rate of convergence needs to be established. These results can be used to tune the ``burn-in'' time $I_{b}$ and the number of intervals $I$ after the chain has sufficiently converged. In~\cite{WellerThesis}, the convergence rates of the Gibbs sampler and Gibbs sampler using slice sampling are simulated to determine trends in $M$, $\sigma_{z}$, and $\sigma_{w}$. The essential result shown is that the convergence rate increases when jitter or additive noise power decrease, or when the oversampling factor increases.

\subsubsection{MSE Performance} 
\begin{figure}[!t]
\centering
\includegraphics[width=3in]{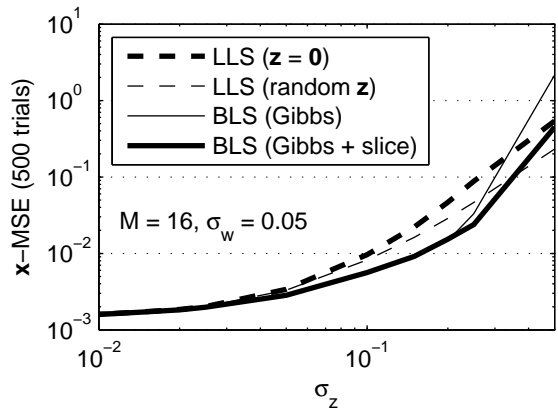}
\caption{Gibbs Sampler Performance: This plot compares both Gibbs sampler algorithms to the random-jitter and no-jitter LLS estimators (in~\eqref{eq:bayesllsest} and~\eqref{eq:bayesllsestnoz}, respectively). All four MSE curves are plotted for $K = 10$ parameters, $M = 16$, and $\sigma_{w} = 0.05$. Additional data can be found in~\cite{WellerThesis}.}
\label{fig:gibbscomplin}
\end{figure}

The MSE performance of both Gibbs samplers is determined for a wide range of oversampling factors $M$, jitter variance $\sigma_{z}^{2}$, and additive noise variance $\sigma_{w}^{2}$. Like the EM algorithm, the Gibbs samplers are expected to outperform both linear estimators when the jitter dominates the additive noise ($\sigma_{z} \gg \sigma_{w}$) and when the jitter is not so large that the Gibbs sampling methods become inaccurate for the given number of samples. The Gibbs sampler using slice sampling is also expected to be more accurate for large $\sigma_{z}$, since rejection sampling fails (the acceptance probability becomes too small). Figure~\ref{fig:gibbscomplin} displays all these expected behaviors, for the example of $M = 16$ and $\sigma_{w} = 0.05$.

\begin{figure}[!t]
\centering
\includegraphics[width=3in]{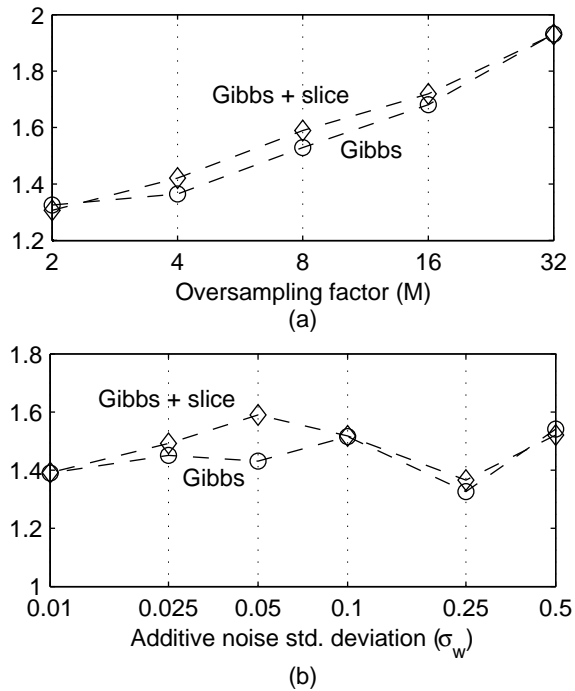}
\caption{Gibbs Sampler Performance: These graphs show the maximum factor of improvement in jitter tolerance, measured by $\sigma_{z}$, achievable by both Gibbs samplers. Holding $\sigma_{w} = 0.1$ fixed, (a) shows the trend in maximum improvement as $M$ increases, and (b) shows the trend in maximum improvement as $\sigma_{w}$ increases holding $M = 8$ fixed.}
\label{fig:gibbssigmazimprovement}
\end{figure}

To summarize the performance of the Gibbs samplers, the maximum factor of improvement in $\sigma_{z}$ attainable by using these sampling algorithms is plotted against increasing $M$ and $\sigma_{w}$ in Figure~\ref{fig:gibbssigmazimprovement}. These plots portray the same trends as observed for the EM algorithm; however, these algorithms appear to handle the high $\sigma_{w}$ case better than the EM algorithm does.

\section{Conclusion}\label{sec:conclusion}

The results of the previous section are very encouraging from a power-consumption standpoint. A maximum improvement of between $1.4$ to $2$ times the jitter translates to a two-to-fourfold decrease in power consumption, according to~\eqref{eq:speedpoweraccuracytradeoff}. To put the magnitude of such an improvement in context, consider the digital baseband processor for ultra-wideband communication in~\cite{Blazquez05}. This processor incorporates an ADC and a PLL, which consume $86$ mW and $45$ mW, respectively, out of a $271$ mW budget for the chip. Reducing by a factor of two the power consumed by the ADC alone would decrease the total power consumption of the chip by almost sixteen percent. 

While effective, both the EM algorithm and the Gibbs samplers are computationally expensive. One benefit of digital post-processing is that these algorithms can be performed off-chip, on a computer or other system with less limited computational resources. For real-time on-chip applications, Kalman-filter-like versions of the EM algorithm or Gibbs sampler would be more practical; this extension is a topic for further investigation. Related to real-time processing is developing streaming algorithms for the infinite-dimensional case, extending this work for general real-time sampling systems. Another future direction involves modifying these algorithms for correlated or periodic jitter.

Sampling jitter mitigation is actually just one application of these new algorithms. In the frequency domain, jitter maps to uncertainty in frequency; using algorithms such as these should produce more reliable Fourier transforms for systems like spectrum analyzers. In higher dimensions, timing noise becomes location jitter in images or video. Greater tolerance of the locations of pixels in images would allow camera manufacturers to place smaller pixels closer together, enabling higher quality images to be acquired using conventional photosensor technology. This paper shows that significant improvements over the best \emph{linear} post-processing are possible; thus, further work may impact these and other applications.


%

%
%
%
\section*{Acknowledgment}

The authors thank V. Y. F. Tan for valuable discussions on Gibbs sampling and J. Kusuma for asking stimulating questions about sampling and applications of jitter mitigation. The authors also thank Z. Zvonar at Analog Devices and G. Frantz at Texas Instruments for their insights and support.

\ifCLASSOPTIONcaptionsoff
  \newpage
\fi

\ifCLASSOPTIONdraftcls
\else
	\IEEEtriggeratref{21}
\fi


\bibliographystyle{IEEEtran}
\bibliography{TSP-jittercompensation}
\end{document}